\pdfoutput=1
\documentclass[conference]{IEEEtran}
\usepackage{textcomp}
\usepackage[T1]{fontenc}
\usepackage{amsmath}
\usepackage{graphicx}
\usepackage{listings}
\usepackage{xcolor}
\usepackage{float}
\usepackage{url}
\usepackage{balance}
\usepackage{epstopdf}
\usepackage{subfig}
\usepackage{cprotect}
\usepackage{framed}
\usepackage{tikz}

\lstdefinestyle{customc}{
  belowcaptionskip=1\baselineskip,
  breaklines=true,
  xleftmargin=\parindent,
  language=C,
  showstringspaces=false,
  basicstyle=\footnotesize\ttfamily,
  keywordstyle=\bfseries\color{green!40!black},
  commentstyle=\itshape\color{purple!40!black},
  identifierstyle=\color{blue},
  stringstyle=\color{orange},
}

\def\note#1{}

\newcommand{\urlwofont}[1]{\urlstyle{same}\url{#1}}

\newcounter{ecount}

\def\BEGINITEMIZE{\begin{itemize}}
\def\BEGINENUMERATE{\begin{enumerate}}
\def\ENDITEMIZE{\end{itemize}}
\def\ENDENUMERATE{\end{enumerate}}

\newcommand\copyrighttext{
    \footnotesize \textcopyright 2017 IEEE. Personal use of this material is permitted.
    Permission from IEEE must be obtained for all other uses, in any current or future
    media, including reprinting/republishing this material for advertising or promotional
    purposes, creating new collective works, for resale or redistribution to servers or
    lists, or reuse of any copyrighted component of this work in other works.
    DOI: 10.1109/ICAC.2017.24 (\url{https://dx.doi.org/10.1109/ICAC.2017.24})
  }
\newcommand\copyrightnotice{
\begin{tikzpicture}[remember picture,overlay]
\node[anchor=south,yshift=10pt] at (current page.south) {\fbox{\parbox{\dimexpr\textwidth-\fboxsep-\fboxrule\relax}{\copyrighttext}}};
\end{tikzpicture}%
}

\begin{document}


\title{Multiverse: Easy Conversion of Runtime Systems into OS Kernels
  via Automatic Hybridization}


\author{\IEEEauthorblockN{Kyle C. Hale}
    \IEEEauthorblockA{Department of Computer Science\\
        Illinois Institute of Technology\\
    khale@cs.iit.edu}
    \and
    \IEEEauthorblockN{Conor Hetland, Peter Dinda}
    \IEEEauthorblockA{Department of Electrical Engineering and Computer
        Science\\
        Northwestern University\\
    ch@u.northwestern.edu, pdinda@northwestern.edu}
}

\maketitle
\copyrightnotice

\begin{abstract}
The hybrid runtime (HRT) model offers a path towards high performance and
efficiency.  By integrating the OS kernel, runtime, and application,
an HRT allows the runtime developer to leverage the full feature set of the
hardware and specialize OS services to the runtime's needs.  However,
conforming to the HRT model currently requires a port of the runtime to the
kernel level, for example to the Nautilus kernel framework, and this requires
knowledge of kernel internals.  In response, we developed Multiverse, a system
that bridges the gap between a built-from-scratch HRT and a legacy runtime
system. Multiverse allows unmodified applications and runtimes to be brought
into the HRT model without any porting effort whatsoever by splitting the
execution of the application between the domains of a legacy OS and an HRT
environment.  We describe the design and implementation of Multiverse and
illustrate its capabilities using the massive, widely-used Racket runtime system.
\end{abstract}




\section{Introduction}
\label{sec:intro}


Runtime systems can gain significant benefits from executing in a
tailored software environment, such as our Hybrid Runtime
(HRT)~\cite{HALE:2015:NAUTILUS}. In an HRT, a light-weight kernel
framework (called an AeroKernel), a runtime, and an application
coalesce into a single kernel-level entity.  The OS {\em is} this
composite of the application, runtime, and AeroKernel.  As such, the
runtime and application enjoy a base platform of fully privileged
access to the underlying hardware, and can also construct
task-appropriate abstractions on top of this base, instead of being
limited to abstractions provided by a commodity OS.  These
capabilities demonstrably enhance performance, scalability, and
efficiency, particularly for parallel runtime systems running on
current NUMA server hardware and next generation high core-count
multicore processors.  The capabilities also enable forms of
adaptation, both during the design process and during execution, that
are simply not available to user-level systems.

An AeroKernel facilitates the creation of HRTs by providing core kernel
functionality and optional mechanisms whose interfaces are geared to
user-level developers instead of kernel developers.  An AeroKernel
helps ease the migration of user-level code to kernel-level.   
The motivation for an AeroKernel draws from
the reliable performance of light-weight
kernels~\cite{LANGE-PALACIOS-KITTEN-IPDPS10, KELLY:2005:CATAMOUNT,
GIAMPAPA:2010:CNK}, the philosophy regarding kernel abstractions of
Exokernel~\cite{ENGLER:1995:EXOKERNEL}, new techniques and ideas developed in
multi-core OS research~\cite{LIU:2009:TESSELATION, FEITELSON:1992:GANG}, and
the simplicity of other experimental OSes from previous
decades~\cite{HUNT:2007:SINGULARITY, ROSCOE:1994:NEMESIS}.  In this paper, we
leverage the Nautilus AeroKernel~\cite{HALE-NAUT-HVM-VEE16}, which we describe in more detail in
Section~\ref{sec:hrthvm}. 

Prior to the work and system we describe here, the implementation of
an HRT consisted entirely of manual processes.  HRT developers needed
first to extend an AeroKernel framework such as Nautilus with the
functionality the runtime needed.  The HRT developers would then port
the runtime to this AeroKernel manually. While a manual port can produce
the highest performance gains, it requires an intimate familiarity with the
runtime system's functional requirements, which may not be
obvious. These requirements must then be implemented in the AeroKernel
layer and the AeroKernel and runtime combined.  This requires a deep 
understanding of kernel development.   This manual process is also
iterative:  the developer adds AeroKernel functionality until the runtime works
correctly. The end result might be that the AeroKernel interfaces
support a small subset of POSIX, or that the runtime developer
replaces such functionality with custom interfaces.  

While such a development model {\em is} tractable
(\cite{HALE-NAUT-HVM-VEE16} gives three examples), it represents a
substantial barrier to entry to creating HRTs, which we seek here to
lower.  The manual porting method is {\em additive} in its nature. We
must add functionality until we arrive at a working system. A more
expedient method would allow us to {\em start} with a working HRT
produced by an automatic process, and then incrementally extend it and
specialize it to enhance its performance.

The Multiverse system we describe in this paper supports such a
method using a technique called {\em automatic hybridization} to
create a working HRT from an existing, unmodified runtime and
application.  With Multiverse, runtime developers can take an
incremental path towards adapting their systems to run in the HRT
model.  From the user's perspective, a hybridized runtime and
application behaves the same as the original.  It can be run from a Linux
command line and interact with the user just like any other
executable.  But internally, it executes in kernel mode as an HRT.

While in this paper we present one instance of an AeroKernel, Multiverse can
work with any AeroKernel (or specialized OS kernel). Such pairings could enable
new forms of adaptive computing, both in datacenter and HPC environments. For
example, hybridization decisions could be made at runtime to merge an
application or runtime system with the most suitable specialized OS kernel in
response to, e.g., application characteristics or phases, hardware capabilities,
or energy constraints. In this sense, Multiverse would behave as a kind of
dynamic linker for applications/runtimes and AeroKernels in that it would
support runtime binding of application functionality to specialized kernel
services.

By enabling an existing Linux program to run partially in kernel mode as an
HRT, Multiverse has the potential to expose additional adaptation mechanisms,
enable additional adaptation policies, and allow both to be incrementally added
to the program. That is, Multiverse expands the range of autonomic computing
possible within the program. For example, the initial program could be extended
over time to have direct control over the hardware paging and timer mechanisms
when it is running in HRT mode.  A specialized policy could then be added to
the program to drive these mechanisms, for example to lower TLB miss rates by
making use of program-specific information, and to achieve program-specific
desirable scheduling behavior, such as real-time behavior.  

Although our general focus has been on supporting parallel programs and
runtimes, we note that a related concept to HRTs, namely
Unikernels~\cite{MADHAVAPEDDY:2013:UNIKERNELS}, has garnered considerable
interest in the cloud/datacenter computing space.  A similar premise to that
given above applies in this space:   A Multiverse-like approach expands the
range of autonomic computing possible.

%

Multiverse bridges a specialized HRT with a legacy environment by
borrowing functionality from a legacy OS, such as Linux.  Functions
not provided by the existing AeroKernel are forwarded to another core
that is running the legacy OS, which handles them and returns their
results. The runtime developer can examine traces of these forwarded 
events, identify hot spots in the
legacy interface, and move their implementations (possibly even
changing their interfaces) into the AeroKernel. The porting process
with Multiverse is {\em subtractive} in that a developer iteratively
removes dependencies on the legacy OS.   At the same time, the
developer can take advantage of the kernel-level environment of the
HRT. 

To demonstrate the capabilities of Multiverse, we automatically
hybridize the Racket runtime system.  Racket has a complex, JIT-based
runtime system with garbage collection and makes extensive use of the
Linux system call interface, memory protection mechanisms, and 
external libraries.  Hybridized Racket executes in kernel mode as an
HRT, and yet the user sees precisely the same interface (an
interactive REPL environment, for example) as out-of-the-box Racket.

Our contributions in this paper are as follows:
\BEGINITEMIZE
\item We introduce the concept of {\em automatic hybridization} for
  transforming runtime systems and their applications into HRTs,
  enabling them to run in kernel mode with full access to hardware
  features and the ability to adapt the kernel to their needs.
\item We describe the design of Multiverse, an implementation of
  automatic hybridization that combines  compile-time, link-time,
  run-time, and virtualization-based techniques. 
\item We demonstrate automatic hybridization with Multiverse by
  transforming the Racket runtime into an HRT.
\item We evaluate the performance of Multiverse.
\ENDITEMIZE

Multiverse will be made publicly available.

\section{HRT and HVM}
\label{sec:hrthvm}

Multiverse builds on the previously described Nautilus AeroKernel and
Hybrid Virtual
Machine~\cite{HALE:2015:NAUTILUS,HALE-NAUT-HVM-VEE16}, which 
were developed to support the hybrid runtime (HRT) model.  We describe the
key salient findings and components here.

The core premise of the HRT model is that by moving the 
runtime (and its application) to the kernel level, we enable the
runtime developer to leverage all hardware features (including
privileged features), and to specialize kernel features specifically
for the runtime's needs.  These capabilities in turn allow for greater
performance or efficiency than possible at user-level.  The Nautilus
AeroKernel facilitates doing exactly this.  Nautilus runs on bare metal or
under virtualization on x64 machines and the Intel Xeon Phi.  It is open
source (MIT license) and publicly available.

Three runtimes have been hand-ported to Nautilus, namely
Legion~\cite{BAUER:2012:LEGION}, the NESL VCODE
interpreter~\cite{BLELLOCH-NESL-JPDC94}, and the runtime of a home-grown nested
data parallel language.  Using the HPCG (High Performance Conjugate Gradient)
benchmark~\cite{DONGARRA-HPCG-TOWARDS-13,HEROUX-HPCG-TECH-13} developed by
Sandia National Labs and ported to Legion by Los Alamos National Labs, speedups over Linux of up to 20\% for the Intel Xeon Phi, and up
to 40\% for a 4-socket, 64-core x64 AMD Opteron 6272 machine were measured. Nautilus
provides basic primitives, for example thread creation and events, that
outperform Linux by orders of magnitude because they are designed to support runtimes in lieu of general-purpose computing, and because
there are no kernel/user boundaries to cross.  This combined with hardware and
software capabilities available only in kernel mode, for example complete
interrupt control and runtime-specific scheduling, leads to these performance
gains in applications.
 
Multiverse also builds on the Hybrid Virtual Machine (HVM), 
an extension to the open source (BSD license) Palacios
VMM~\cite{LANGE-PALACIOS-KITTEN-IPDPS10}, and available in its
repository.  HVM allows for the
creation of a VM whose memory, cores, and interrupt logic are
segregated so that one VM simultaneously runs two operating
systems, the ``Regular Operating System'' (ROS) (e.g., Linux) and an
HRT-based OS (e.g., Nautilus).  The ROS runs on a partition of the
cores and can only see and touch the ROS cores and the ROS subset of
physical memory.  In contrast, the HRT, while only allowed to run on its own
distinct partition of the cores, has full access to all the memory,
cores, and interrupt logic of the entire VM.  The HVM
design and the Nautilus design are closely coupled, with the result being
that HRTs based on Nautilus can execute with negligible virtualization
overheads on Palacios.   
The ROS and HRT can be booted and rebooted independently, the latter with a latency 
comparable to a \verb.fork()./\verb.exec(). in Linux.  That is, a boot
of a Nautilus-based HRT is comparable to a Linux process creation.

\section{Multiverse}
\label{sec:multiverse}

We designed the Multiverse system to support automatic hybridization
of existing runtimes and applications that run in user-level on Linux
platforms.  

\subsection{Perspectives}

Multiverse's goal is to ease the path for developers
transforming a runtime into an HRT.  We seek to make the system look
like a compilation option from the developer's perspective.
That is, to the greatest extent possible, the HRT is a compilation
target.  Compiling to an HRT results in an executable that is a
``fat binary'' containing additional code and data that enables
kernel-mode execution in an environment that supports it.  An
HVM-enabled virtual machine on Palacios is the first such environment.  The
developer can extend this incrementally; Multiverse facilitates
a path for runtime and application developers to explore how to
specialize their HRT to the full hardware feature set and the
extensible kernel environment of the AeroKernel.

From the user's perspective, the executable behaves as if it were
compiled for a user-level Linux environment.   The user sees no
difference between HRT and user-level execution. 
 
\subsection{Techniques}
The Multiverse system relies on three key techniques: split execution,
event channels, and state
superpositions.  

\subsubsection*{Split execution}
\label{sec:split-execution}

In Multiverse, a runtime and its application begin their execution in
the ROS as an ordinary Linux process. 
Through a well-defined interface discussed in Section~\ref{sec:usage}, the
runtime on the ROS side can spawn an execution context in the HRT. At this
point, Multiverse splits its execution into two components, each running in
a different context; one executes in the ROS and the other in the HRT;
there is now a Linux process and a kernel.  The
semantics of these execution contexts differ from traditional threads depending
on their characteristics.
We discuss these differences in Section~\ref{sec:impl}.  In the
current implementation, the context on the ROS side comprises a Linux thread, the
context on the HRT side comprises an AeroKernel thread, and we refer to them
collectively as an {\em execution group}. While execution groups in our current
system consist of threads in different OSes, this need not be true in general.
The context on the HRT side executes until it triggers a fault, a system call,
or other event. The execution group then converges on this event, with each
side participating in a protocol for requesting events and receiving results.
This protocol exchange occurs in the context of HVM event channels, which we discuss
below.

\begin{figure}
\centering
\includegraphics[width=.8\columnwidth]{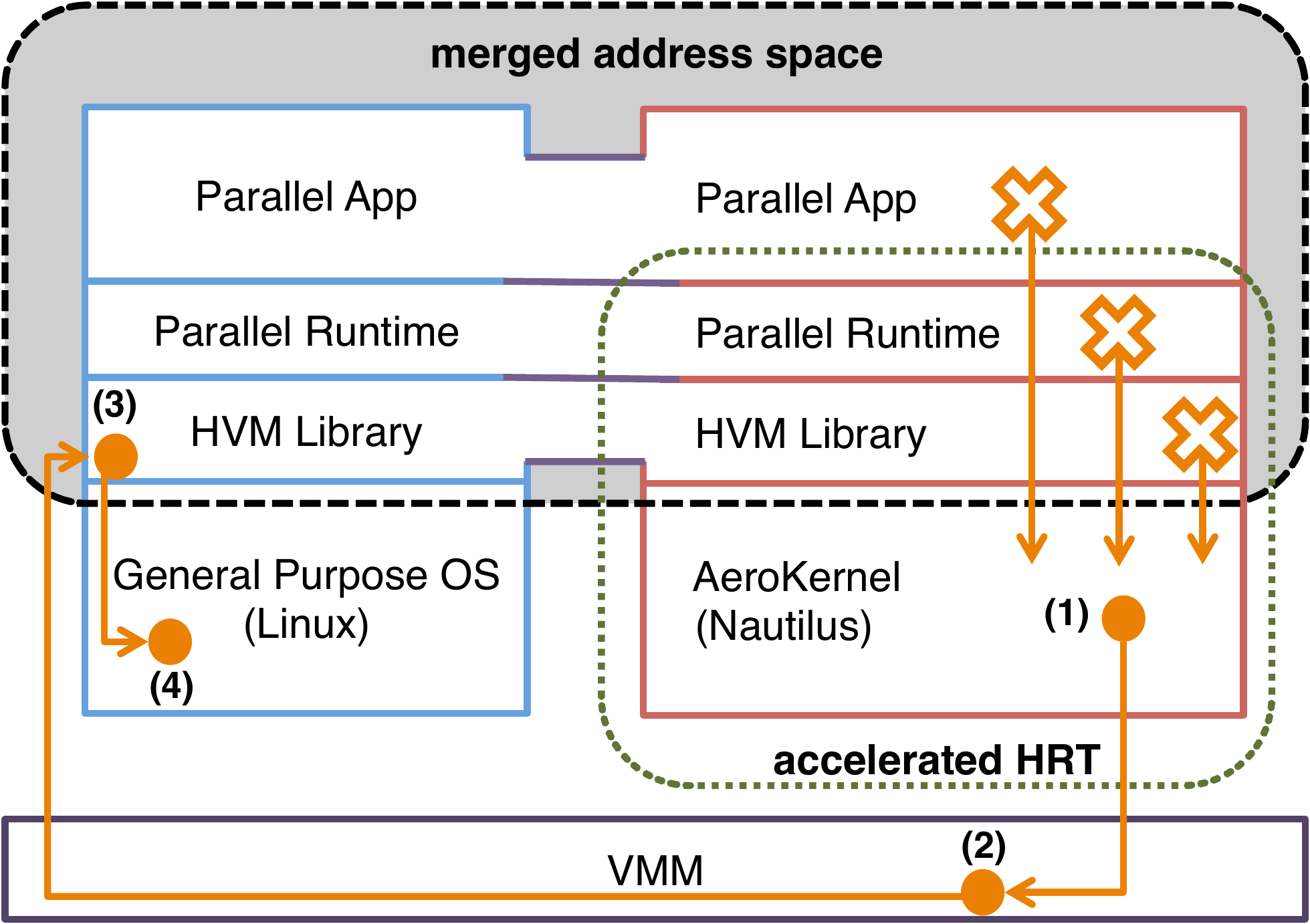}
\caption{Split execution in Multiverse.}
\label{fig:mverse-operation}
\end{figure}

Figure~\ref{fig:mverse-operation} illustrates the split execution of
Multiverse for a ROS/HRT execution group. At this point, the ROS has
already made a request to create a new context in the HRT, e.g. through
an asynchronous function invocation.  When the HRT thread begins executing in the
HRT side, exceptional events, such as page faults, system calls, and other
exceptions vector to stub handlers in the AeroKernel (1). The AeroKernel
redirects these events through an event channel (2) to request handling in the
ROS. The VMM then injects these into the originating ROS thread,
which can take action on them directly (3). For example, in the case of a page
fault that occurs in the ROS portion of the virtual address space, the HVM
library replicates the access, which will cause the same exception to
occur on the ROS core. The ROS will then handle it as it would normally. 
For events that need direct handling by the ROS kernel, such as system
calls, the HVM library can forward them (4).

\subsubsection*{Event channels}
\label{sec:event-channels}

When the HRT needs functionality that the ROS implements, 
access to that functionality occurs over {\em event channels}, 
event-based, VMM-controlled communication channels between the two contexts. 
The VMM only expects that the execution group adheres to a strict protocol
for event requests and completion. 

\begin{figure}
\centerline{
\begin{tabular}{|lll|}
\hline
Item & Cycles & Time \\
\hline \hline
Address Space Merger & $\sim$33 K & 15 $\mu$s \\
Asynchronous Call & $\sim$25 K & 11 $\mu$s \\
Synchronous Call (different socket) & $\sim$1060 &  482 ns\\
Synchronous Call (same socket) & $\sim$790 & 359 ns \\
\hline
\end{tabular}
}
\caption{Round-trip latencies of ROS$\leftrightarrow$HRT interactions.}
\label{fig:hrtroslatency}
\end{figure}

Figure~\ref{fig:hrtroslatency} shows the measured latency of event channels
with the Nautilus AeroKernel performing the role of HRT. Note that these calls
are bounded from below by the latency of hypercalls to the VMM.

\subsubsection*{State superpositions} \label{sec:superpositions}

In order to forego the addition of burdensome complexity to the AeroKernel environment, it
helps to leverage functions in the ROS other than those that lie at a system call
boundary.  This includes functionality implemented in libraries and more opaque
functionality like optimized system calls in the \verb.vdso. and the
\verb.vsyscall. page. To use this functionality, Multiverse can set up
the HRT and ROS to share portions of their address space, in this case the
user-space portion. Aside from the address space merger itself, Multiverse
leverages other state superpositions to support a shared address space,
including superpositions of the ROS GDT and thread-local storage state.

In principle, we could superimpose any piece of state visible to the VMM.  The
ROS or the runtime need not be aware of this state, but the state is
nonetheless necessary for facilitating a simple and approachable usage model.

\begin{figure}
\centering
\includegraphics[width=.8\columnwidth]{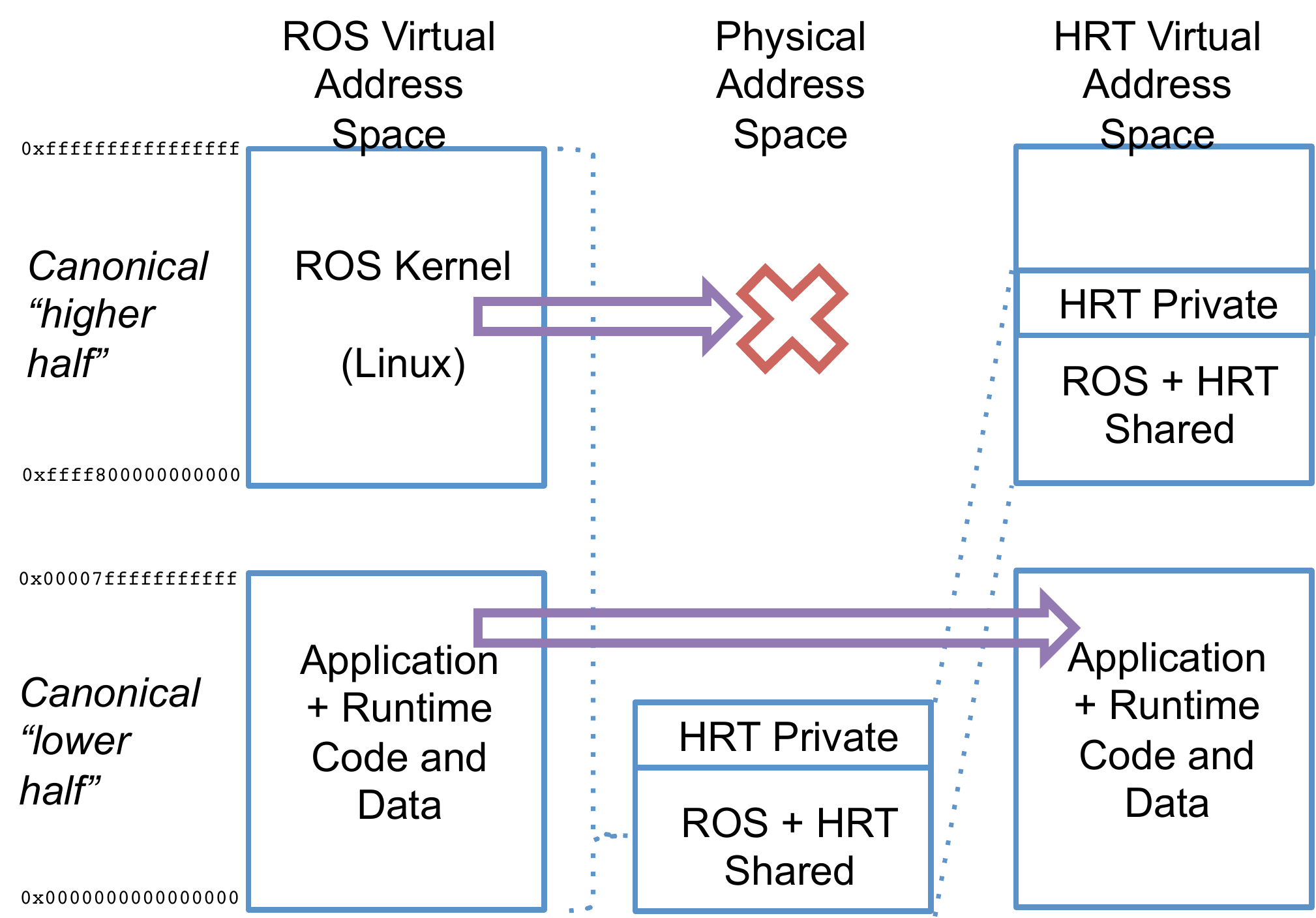}
\caption{Merged address space between ROS and HRT.}
\label{fig:addr-space}
\end{figure}

The superposition we leverage most in Multiverse is a merged address space
between the ROS and the HRT, depicted in Figure~\ref{fig:addr-space}.  The
merged address space allows execution in the HRT without a need for
implementing ROS-compatible functionality.  When a merged address space takes
effect, the HRT can use the same user-mode virtual addresses present in the
ROS.  For example, the runtime in the ROS might load files and
construct a complex pointer-based data structure in memory.  It can then 
invoke a function within its counterpart in the HRT to compute over that data.

\subsection{Usage models}
\label{sec:usage}

The Multiverse system is designed to give maximum flexibility
to application and runtime developers in order to encourage
exploration of the HRT model. While the degree to which a
developer leverages Multiverse can vary, for the purposes of
this paper we classify the usage model into three categories, discussed below.

\subsubsection*{Native}
 In the native model, the application/runtime is ported
to operate fully within the HRT/AeroKernel setting. That is, it does not use any
functionality not exported by the AeroKernel, such as \verb.glibc.
functionality or system calls like \verb.mmap()..  This category allows maximum performance,
but requires more effort, especially in the compilation process. The ROS side
is essentially unnecessary for this usage model, but may be used to simplify
the initiation of HRT execution (e.g. requesting an HRT boot).   The
native model is also native in another sense:  it can execute on
bare metal without any virtualization support.  

\subsubsection*{ Accelerator} 
In this model, the app/runtime developer leverages both 
legacy (e.g. Linux) functionality and AeroKernel functionality. This requires
less effort, but allows the developer to explore some of the benefits of 
running their code in an HRT. Linux functionality is enabled by the merged
address space discussed previously, but the developer
can also leverage AeroKernel functions.

\begin{figure}[h] 
\centering
\begin{tabular}{c}
\begin{lstlisting}[style=customc]
static void*
routine (void * in) {
    void * ret = aerokernel_func();
    printf("Result = %d\n", ret);
}

int main (int argc, char ** argv) {
    hrt_invoke_func(routine);
    return 0;
}
\end{lstlisting}
\end{tabular}
\caption{Example user code adhering to the accelerator model.}
\label{fig:accel-code}
\end{figure}

Figure~\ref{fig:accel-code} shows a small example of code that will create
a new HRT thread and use event channels and state superposition to execute to
completion.  Runtime initialization is opaque to the user, much like
C runtime initialization code. When the program invokes the
\verb.hrt_invoke_func(). call, the Multiverse runtime will make a request to
the HVM to run \verb.routine().  in a new thread on the HRT core. Notice how
this new thread can call an AeroKernel function directly, and then use the
standard \verb.printf(). routine to print its result.  This \verb.printf. call
relies both on a state superposition (merged address space) for the function call
linkage to be valid, and on event channels, which will be used when the C library
code invokes a system call (e.g. \verb.write().).

\subsubsection*{Incremental}
 The application/runtime executes in the HRT context, but does not
initially leverage AeroKernel functionality. Benefits are immediately limited to aspects of the HRT
{\em environment}. However, the developer need only recompile their application
to explore this model. {\bf In the incremental model, the path to
  converting a runtime and its application into a
kernel is straightforward.}  Instead of raising an explicit HRT thread creation
request, Multiverse creates a new thread in the HRT corresponding to the program's
\verb.main(). routine. The Incremental model also allows parallelism, as legacy threading
functionality automatically maps to the corresponding AeroKernel functionality
with semantics matching those used in pthreads.   The developer can
then incrementally expand their usage of hardware- and
AeroKernel-specific features.

While the accelerator and incremental usage models rely on the HVM
virtualized environment of Palacios, it is important to note that they
could also be built on physical
partitioning~\cite{OUYANG-PISCES-HPDC15} as well.  At its core, HVM
provides to Multiverse a resource partitioning, the ability to boot
multiple kernels simultaneously on distinct partitions, and the
ability for these kernels to share memory and communicate.

\subsection{Function overrides}
\label{sec:aeroover}

One way a developer can enhance a generated HRT is through {\em
function overrides}.  The AeroKernel can implement functionality that conforms to
the interface of, for example, a standard library function, but that may be more
efficient or better suited to the HRT environment. This technique allows users
to get some of the benefits of the accelerator model without any explicit
porting effort. However, it is up to the AeroKernel developer to ensure that
the interface semantics and any usage of global data make sense when using
these function overrides. Function overrides are specified in a simple
configuration file that is discussed in Section~\ref{sec:impl}.

\begin{figure}[h]
\begin{tabular}{c}
\begin{lstlisting}[style=customc]
static void*
routine (void * in) {
    void * ret = aerokernel_func();
    printf("Result = %d\n", ret);
}

int main (int argc, char ** argv) {
    pthread_t t;
    pthread_create(&t, NULL, routine, NULL);
    pthread_join(t, NULL);
    return 0;
}
\end{lstlisting}
\end{tabular}
\caption{Example of user code using overrides.}
\label{fig:accel-code-override}
\end{figure}

Figure~\ref{fig:accel-code-override} shows the same code from Figure~\ref{fig:accel-code} 
using function overrides. Here the AeroKernel developer has overridden the standard pthreads
routines so that \verb.pthread_create(). will create a new HRT thread in the same way
that \verb.hrt_invoke_func(). did in the previous example.

\subsection{Toolchain}
\label{sec:toolchain}

The Multiverse toolchain consists of two main components, the runtime system
code and the build setup.  The build setup consists of build tools,
configuration files, and an AeroKernel binary provided by the AeroKernel
developer. To leverage Multiverse, a user integrates their application or
runtime with the provided Makefile and rebuilds it. This will result in the
compilation of the AeroKernel components necessary for HRT operation and the
Multiverse runtime system, including function overrides, exit and signal
handlers, and initialization code, into the user program.

\section{Implementation}
\label{sec:impl}
We now discuss the implementation of Multiverse. This includes the components
that are
automatically compiled and linked into the application's address space at build
time and the parts of Nautilus and the HVM that support event channels and
state superpositions. Unless otherwise stated, we assume the Incremental usage
model discussed in Section~\ref{sec:usage}.

\subsection{Multiverse runtime initialization}
\label{sec:boot}

As mentioned in Section~\ref{sec:multiverse}, a new HRT thread must be created
from the ROS side (the originating ROS thread). This, however, requires an
AeroKernel present on the requested core to create that thread. The
runtime component (which includes the user-level HVM library) is in charge of booting an
AeroKernel on all required HRT cores during program startup. They can either be booted
on demand or at application startup. We use the latter in our current setup.

Our toolchain inserts program initialization hooks before the
program's \verb.main().  function, which carry out runtime
initialization, including:
\BEGINITEMIZE
\item Registering ROS signal handlers
\item Hooking process exit for HRT shutdown
\item AeroKernel function linkage
\item AeroKernel image installation in the HRT
\item AeroKernel boot
\item Merging ROS and HRT address spaces
\ENDITEMIZE

\begin{figure}
\centering
\includegraphics[width=.8\columnwidth]{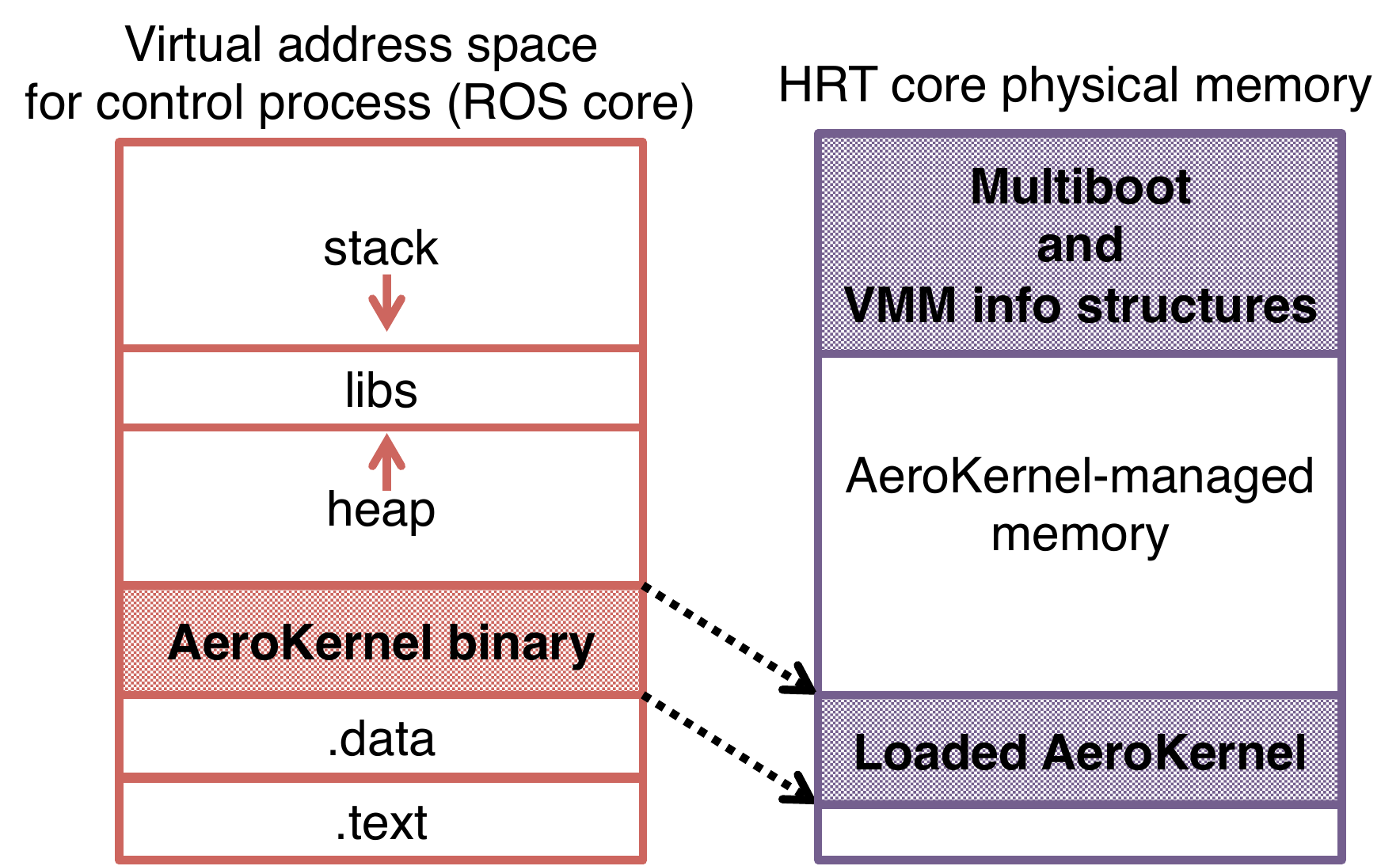}
\caption{AeroKernel boot process.}
\label{fig:hrt-boot}
\end{figure}

\subsubsection*{AeroKernel Boot} 
Our toolchain embeds an AeroKernel binary into the ROS
program's ELF binary. This is the image to be installed in the HRT. At program
startup, the Multiverse runtime component parses this embedded AeroKernel
binary and sends a request to the HVM asking that it be installed in physical
memory, as shown in Figure~\ref{fig:hrt-boot}.  Multiverse then requests
the AeroKernel be booted on the HRT cores. The boot process brings the AeroKernel up into an event
loop that waits for HRT thread creation requests.

The above initialization tasks are opaque to the user.


\subsection{Execution model}

To implement split execution, we rely on HVM's ability to forward 
requests from the ROS core to the HRT, along with event channels and
merged address spaces. 

The runtime developer can use two mechanisms to create HRT 
threads, as discussed in Section~\ref{sec:usage}. Furthermore, 
two types of 
threads are possible on the HRT side: top-level threads and nested
threads. Top-level threads are explicitly created by the ROS.
A top-level HRT thread can create its own child threads as well; we classify these
as nested threads. The semantics of the two thread types differ slightly
in their operation. Nested threads resemble pure AeroKernel threads, but
their execution can proceed in the context of the ROS user address
space. Top-level threads require extra semantics in the HRT and in the
Multiverse component linked with the ROS application. 

\begin{figure}
\centering
\includegraphics[width=0.6\columnwidth]{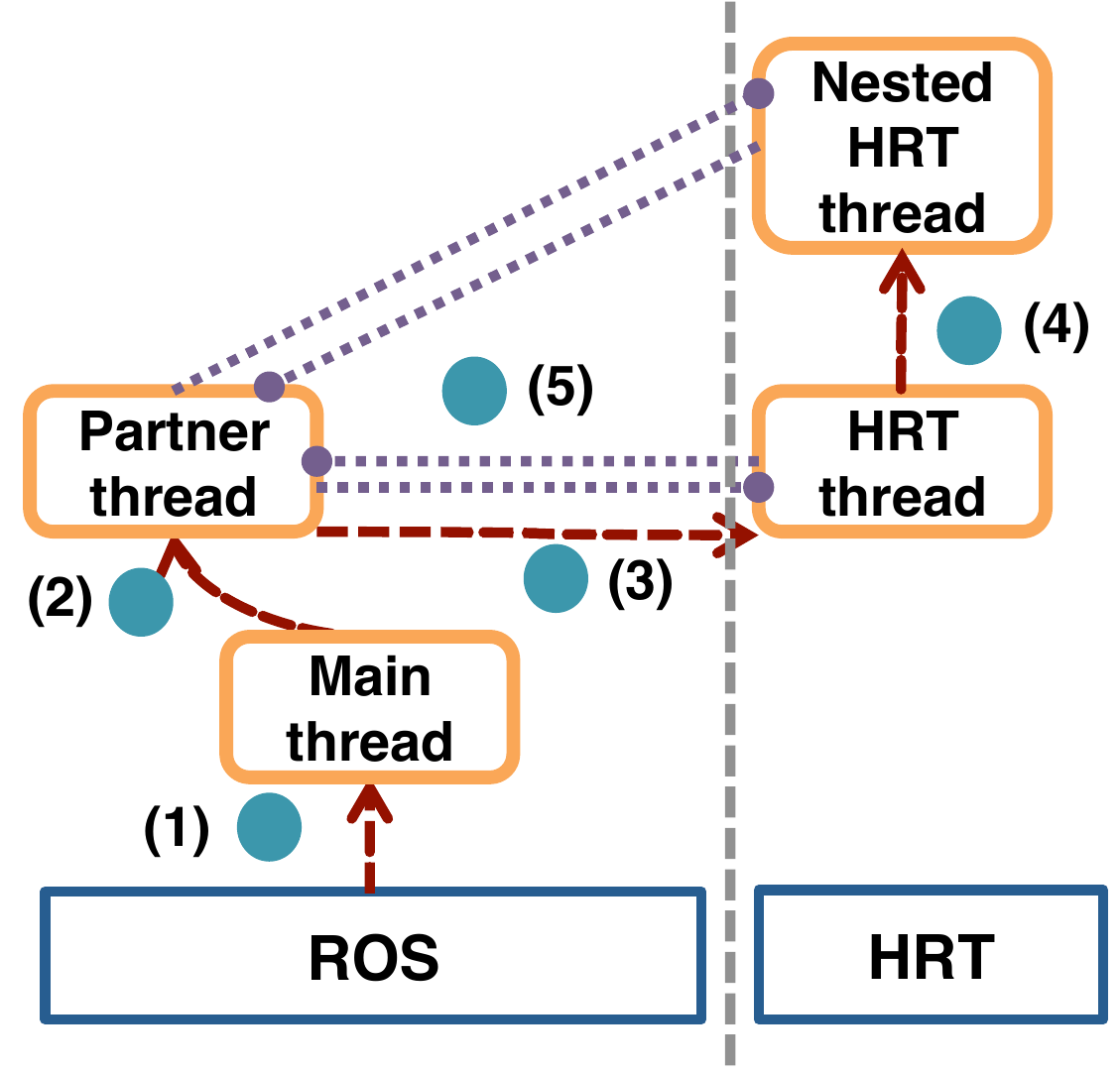}
\caption{Interactions within an execution group.}
\label{fig:hrt-threads}
\end{figure}

\subsubsection*{Threads}
Multiverse pairs each top-level HRT thread with a {\em partner}
thread that executes in the ROS. This thread has two purposes.
First, it allows us to preserve join semantics. Second, it gives us the proper
thread context in the ROS to initiate a state superposition for the HRT.
Figure~\ref{fig:hrt-threads} depicts the creation of HRT threads and their
interaction with the ROS. First, in (1) the main thread is created in the ROS.
It sets up the runtime environment for Multiverse.  When the runtime
system creates a thread, e.g. with \verb.pthread_create(). or with
\verb.hrt_invoke_func()., Multiverse creates a corresponding partner thread
that executes in the ROS (2). It is the duty of the partner thread to allocate a
ROS-side stack for a new HRT thread and then
invoke the HVM to request a thread creation in the HRT using that stack (3). When the partner
creates the HRT thread, it also conveys information to initiate a state
superposition that mirrors the ROS-side GDT and ROS-side architectural state
corresponding to thread-local storage (primarily the \verb.%fs. register). The
HRT thread can then create as many nested HRT threads as it desires (4). Both
top-level HRT threads and nested HRT threads raise events to the ROS through
event channels with the top-level HRT thread's partner thread acting as
the communication end-point (5).

As is typical in threading models, the main thread can wait for HRT
threads to finish by using \verb$join()$ semantics, where the joining thread
blocks until the child exits. While in theory we could implement the ability to
join an HRT thread directly, it would add complexity to both the HRT and the
ROS component of Multiverse.  Instead, we chose to allow the main thread to
join a partner thread directly and provide the guarantee that a partner thread
will not exit until its corresponding HRT thread exits on the remote core. When
an HRT thread exits, it signals the ROS of the exit event. When Multiverse
creates an HRT thread, it keeps track of the Nautilus thread data (sent from
the remote core after creation succeeds), which it uses to build a mapping from
HRT threads to partner threads. The thread exit signal handler in the ROS flips
a bit in the appropriate partner thread's data structure notifying it of the
HRT thread completion. The partner can then initiate its cleanup routines and
exit, at which point the main thread will be unblocked from its initial
\verb$join()$.

%
%

\subsubsection*{Function overrides}
In Section~\ref{sec:usage} we described how a developer can use function 
overrides to select AeroKernel functionality over default ROS functionality.
The Multiverse runtime component enforces default overrides that 
interpose on \verb.pthread. function calls. All function overrides operate
using function wrappers. For simple function wrappers, the AeroKernel 
developer can simply make an addition to a configuration file included
in the Multiverse toolchain that specifies the function's attributes 
and argument mappings between the legacy function and the AeroKernel
variant. This configuration file then allows Multiverse to 
automatically generate function wrappers at build time.

When an overridden function is invoked, the wrapper
runs instead, consults a stored mapping to find the symbol name for the
AeroKernel variant, and performs a lookup to find its HRT virtual address.
This symbol lookup currently occurs on every function invocation, so incurs
a non-trivial overhead. A symbol cache, much like that used in the ELF standard,
could easily be added to improve lookup times. When the address of the 
AeroKernel override is resolved, the wrapper invokes the function directly
(since it is already executing in the HRT context where it has appropriate page
table mappings for AeroKernel addresses).

\subsection{Event channels}

The HVM model enables the building of essentially any communication mechanism
between two contexts (in our case, the ROS and HRT), and most of
these require no specific support in the HVM.  As a consequence, we minimally
define the {\em basic} communication between the ROS, HRT, and the VMM using
shared physical memory, hypercalls, and interrupts.

The user-level code in the ROS can use hypercalls to sequentially request HRT
reboots, address space mergers (state superpositions), and asynchronous
sequential or parallel function calls.  The VMM handles reboots internally, and
forwards the other two requests to the HRT as special exceptions or interrupts.
Because the VMM and HRT may need to share additional information, they share
a data page in memory.  For a function call request, the page contains
a pointer to the function and its arguments at the start and the return code at
completion.  For an address space merger, the page contains the CR3 of the
calling process.  The HRT indicates to the VMM when it is finished with the
current request via a hypercall.  

After an address space merger, the user-level code in the ROS can also use
a hypercall to initiate synchronous operation with the HRT.  This
hypercall indicates to the HRT a virtual address which will be used
for future synchronization between the HRT and ROS.  They can then use a 
memory-based protocol to communicate, for example to allow the
ROS to invoke functions in the HRT without VMM intervention.    

\subsection{Merged address spaces}

To achieve a merged address space, we leverage the canonical 64-bit address
space model of x64 processors, and its wide use within existing kernels, such
as Linux.   In this model, the virtual address space is split into a ``lower
half'' and a ``higher half'' with a gap in between, the size of which is
implementation dependent.  In a typical process model, e.g., Linux, the lower
half is used for user addresses and the higher half is used for the kernel.     

For an HRT that supports it, the HVM arranges that the physical address space
is identity-mapped into the higher half of the HRT address space.  That is,
within the HRT, the physical address space mapping (including the portion of
the physical address space only the HRT can access) occupies the same portion
of the virtual address space that the ROS kernel occupies, namely the higher
half.  Without a merger, the lower half is unmapped and the HRT runs purely out
of the higher half.  When the ROS side requests a merger, we map the lower half
of the ROS's current process address space into the lower half of the HRT
address space.  

%
%
%

%
%
%
%
%

\subsection{Nautilus additions}

In order to support Multiverse in the Nautilus AeroKernel, we needed to make
several additions to the codebase. Most of these focus on runtime
initialization and correct operation of event channels. When the runtime and
application are executing in the HRT, page faults in the ROS portion of the
virtual address space must be forwarded. We added a check in the page fault
handler to look for ROS virtual addresses and forward them appropriately over
an event channel. 

One issue with our current method of copying a portion of the PML4 on an
address space merger is that we must keep the PML4 synchronized. We must
account for situations in which the ROS changes top-level page table mappings,
even though these changes are rare. We currently handle this
by detecting repeat page faults. Nautilus keeps a per-core variable
keeping track of recent page faults, and matches duplicates. If a duplicate is
found, Nautilus will re-merge the PML4. More clever schemes to
detect this condition are possible, but unnecessary since it does not lie
on the critical path.

For correct operation, Multiverse requires that we catch {\em all} page faults
and forward them to the ROS. If we collect a trace of page faults in
the application running native and under Multiverse, the traces should look
identical. However, because the HRT runs in kernel mode, some paging semantics
(specifically with copy-on-write) change. In default operation, an x86 CPU will only
raise a page fault when writing a read-only page in user-mode. Writes to pages
with the read-only bit while running in ring 0 are allowed to proceed. This issue
manifests itself in the form of mysterious memory corruption, e.g. by writing
to the zero page. Luckily, there is a bit to enforce write faults in ring 0 in
the \verb.cr0. control register.

Before we built Multiverse, Nautilus lacked support for system calls, as the
HRT operates entirely in kernel mode. However, a legacy application will 
leverage a wide range of system calls. To support them, we added a small 
system call stub handler in Nautilus that immediately forwards the system 
call to the ROS over an event channel.

\subsection{Complexity}

\begin{figure}
\centering
\begin{tabular}{|l|r|r|r|r|}
\hline
{\em Component} & \multicolumn{4}{|c|}{{\em SLOC}} \\
\hline
                     &  C   & ASM & Perl & Total \\ \hline
Multiverse runtime   & 2232 & 65  &   0  & 2297\\ \hline
Multiverse toolchain &    0 &  0  & 130  & 130 \\ \hline
Nautilus additions   & 1670 &  0  &   0  & 1670 \\ \hline
HVM additions        & 600  & 38  &   0  & 638 \\ \hline
\hline
{\bf Total}          & {\bf 4502} & {\bf 103}  & {\bf 130} & {\bf 4735} \\ \hline
\end{tabular}
\caption{Source Lines of Code for Multiverse.}
\label{fig:sloc}
\end{figure}

Multiverse development took roughly 5 person months of effort.
Figure~\ref{fig:sloc} shows the amount of code needed to support Multiverse.
The entire system is compact and compartmentalized so that users
can experiment with other AeroKernels or runtime systems with relative ease.
While the codebase is small, much of the time went into careful design of
the execution model and working out idiosyncrasies in the hybridization, 
specifically those dealing with operation in kernel mode.

\section{Evaluation}
\label{sec:eval}

We evaluate Multiverse using a hybridized
Racket runtime system running a set of benchmarks from The Language Benchmark
Game.  We ran all experiments on a Dell PowerEdge 415 with 8GB of RAM and an
8 Core 64-bit x86\_64 AMD Opteron 4122 clocked at 2.2GHz.  Each CPU core
has a single hardware thread with four cores per socket. The host machine has stock
Fedora Linux
2.6.38.6-26.rc1.fc15.x86\_64 installed, and is configured for maximum performance in the BIOS.
Benchmark results are reported as averages of 10 runs.

Experiments in a VM were run on a guest setup which consists
of a simple BusyBox distribution running an unmodified Linux 2.6.38-rc5+ image
with two cores (one core for the HVM and one core for the ROS) and 1 GB of RAM.

\begin{figure*}[t]
\centerline{
\resizebox{\textwidth}{!}{\begin{tabular}{|lrrrrrr|}
\hline
Benchmark & System Calls & Time (User/Sys) (s) & Max Resident Set (Kb) & Page Faults & Context Switches & Forwarded Events\\
\hline
spectral-norm & 23800 & 39.39/0.24 & 182300 & 51452 & 1695 & 75252\\
n-body & 18763 & 41.15/0.19 & 152300 & 45064 & 1430 & 63827\\
fasta-3 & 35115 & 31.28/0.17 & 80492 & 25418 & 1075 & 60533\\
fasta & 29989 & 12.23/0.10 & 43568 & 14956 & 627 & 44945\\
binary-tree-2 & 1260 & 31.98/0.10 & 82072 & 31082 & 491 & 32342\\
mandelbrot-2 & 3667 & 7.76/0.05 & 43600 & 14250 & 291 & 17917\\
fannkuch-redux & 1279 & 2.73/0.01 & 21284 & 5358 & 33 & 6637\\
\hline
\end{tabular}
}}
\caption{System utilization for Racket benchmarks.  A high-level
  language has many low-level interactions with the OS.}
\label{fig:racketbenchmarks}
\end{figure*}


Racket~\cite{plt-tr1,manifesto} is the most widely used Scheme
implementation and has been under continuous development for over 20
years.  It is an open source codebase that is downloaded over 300
times per day.  Recently, support has
been added to Racket for parallelism via
futures~\cite{SWAINE-FUTURES-OOPSLA10} and
places~\cite{TEW-PLACES-DLS-11}.

The Racket runtime, which comprises over 800,000 lines of code, is a good candidate to test Multiverse,
particularly its most complex usage model, the incremental model,
because Racket includes many of the challenging features emblematic of
modern dynamic programming languages that make extensive use of the
Linux ABI, including system calls, memory mapping, processes, threads,
and signals.  These features include complex package management via
the filesystem, shared library-based support for native code, JIT
compilation, tail-call elimination, live variable analysis (using
memory protection), and garbage collection.

Our port of Racket to the HRT model takes the form of an instance of
the Racket engine embedded into a simple C program.  Racket already
provides support for embedding an instance of Racket into C, so it was
straightforward to produce a Racket port under the Multiverse
framework. This port uses a conservative garbage collector, the
SenoraGC, which is more portable and less performant than the default,
precise garbage collector. The port was compiled with GCC 4.6.3.  The
C program launches a pthread that in turn starts the engine.  Combined
with the incremental usage model of Multiverse, the result is that the
existing, unmodified Racket engine executes entirely in kernel mode as
an HRT. 

When compiled and linked for regular Linux, our port provides either a
REPL interactive interface through which the user can type Scheme, or
a command-line batch interface through which the user can execute a
Scheme file (which can include other files).  When compiled and linked
for HRT use, our port behaves identically.



To evaluate the correctness and performance of our port, we tested it
on a series of benchmarks submitted to The Computer Language
Benchmarks Game~\cite{BENCHMARKS-GAME}. We tested on seven different
benchmarks: a garbage collection benchmark (binary-tree-2), a
permutation benchmark (fannkuch), two implementations of a random DNA
sequence generator (fasta and fasta-3), a generation of the Mandelbrot
set (mandelbrot-2), an n-body simulation (n-body), and a spectral norm
algorithm. Figure~\ref{fig:racketbenchmarks} characterizes these
benchmarks from the low-level perspective.  Note that while this is an
implementation of a high-level language, the actual execution of
Racket programs involves many interactions with the operating
system. These exercise Multiverse's system call and fault forwarding
mechanisms. The total number of forwarded events is in the last
column.

\begin{figure}
\centering
\includegraphics[width=\columnwidth]{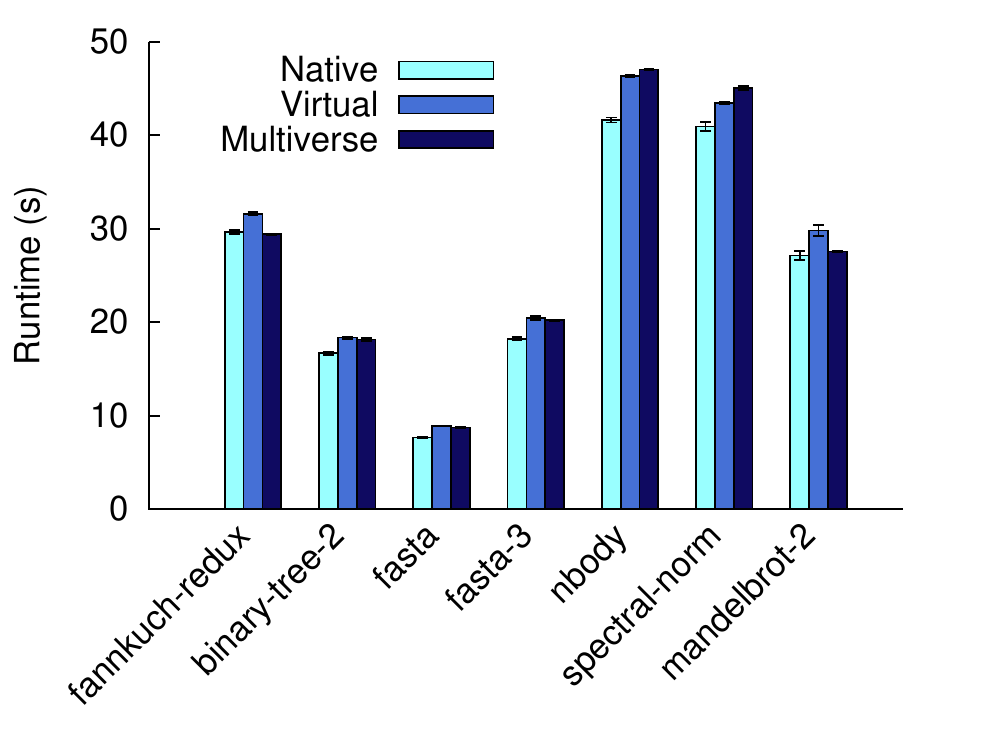}
\caption{Performance of Racket benchmarks running Native, Virtual, and
  in Multiverse.  {\bf With Multiverse, the existing, unmodified
    Racket implementation has been automatically transformed to run
    entirely in  kernel mode, as an HRT, with little to no overhead.}
}
\label{fig:racket-bench}
\end{figure}
 
Figure~\ref{fig:racket-bench} compares the performance of the Racket benchmarks
run natively (note here {\em native} means on bare metal rather than
virtualized) on our hardware, under virtualization, and as an HRT that was
created with Multiverse. Error bars are included, but are barely visible
because these workloads run in a predictable way.   The key takeaway
is that Multiverse performance is on par with native and virtualized
performance---Multiverse let us move, with little to no effort,
the existing, unmodified Racket runtime into kernel mode and run it as
an HRT with little to no overhead. 

The small overhead of the Multiverse case compared to the virtualized and native cases is due to the
frequent interactions, such as those described above, with the Linux ABI.
However, in all but two cases, the hybridized benchmarks actually outperform the
equivalent versions running without Multiverse. This is due to the nature of
the accelerated environment that the HRT provides, which ameliorates the event
channel overheads.  As we expect, the two benchmarks that perform worse under
Multiverse (nbody and spectral-norm) have the most overhead incurred
from event channel interactions. The most frequent interactions in both 
cases are due to page faults. 

While Figure~\ref{fig:racketbenchmarks} tells us the number of
interactions that occur, we now consider the overhead of each 
using microbenchmarks. This estimate will also apply to page faults, since
their forwarding mechanism is identical to system calls.


\begin{figure}
\centering
\includegraphics[width=.6\columnwidth]{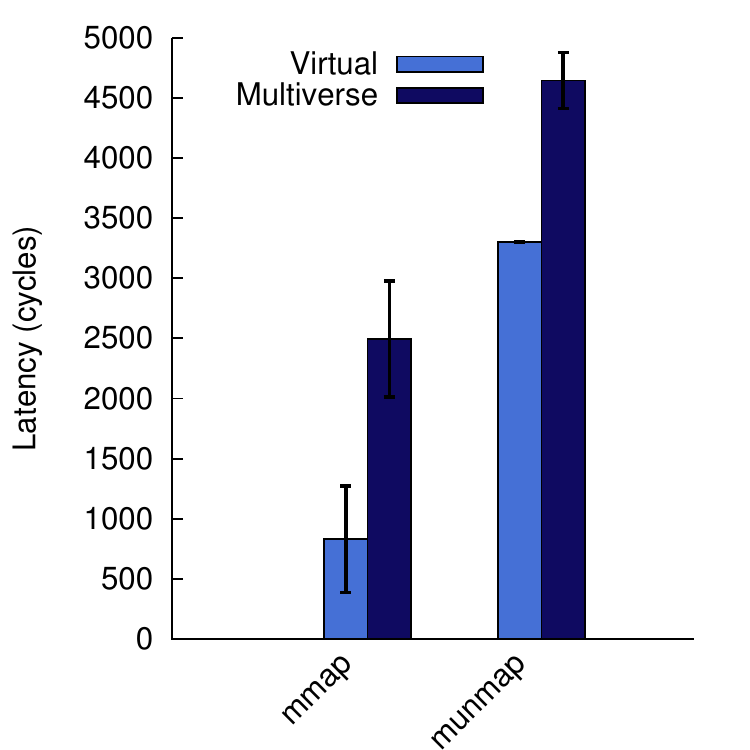}
\cprotect\caption{Multiverse event forwarding overheads demonstrated with \verb|mmap()| and \verb|munmap()| system calls.}
\label{fig:mmap}
\end{figure}

%
%
%

The most frequent system calls used in the Racket runtime (independent of
any benchmark)
are \verb|mmap()| and \verb|munmap()|, and so we focus on these two.
Figure~\ref{fig:mmap} shows microbenchmark results (averages over 100
runs) for these two, comparing virtualized execution and Multiverse.
Note that neither system calls nor page faults involve the VMM in the
virtualized case.  For both system calls, the Multiverse event
forwarding facility adds roughly 1500 cycles of overhead. If
we multiply this number by the number of forwarded events for these
benchmarks listed in Figure~\ref{fig:racketbenchmarks}, we expect
that Multiverse will add about 112 million cycles (51 ms) of overhead
for spectral-norm, and 96 million cycles (43 ms) for nbody. This is
roughly in line with Figure~\ref{fig:racket-bench}. Note that the {\em relative} 
overhead cost for these benchmarks is roughly 0.1\%. While the cost 
{\em per system call} is almost doubled here, such events
are relatively rare.

\begin{figure}
\centering
\includegraphics[width=.9\columnwidth]{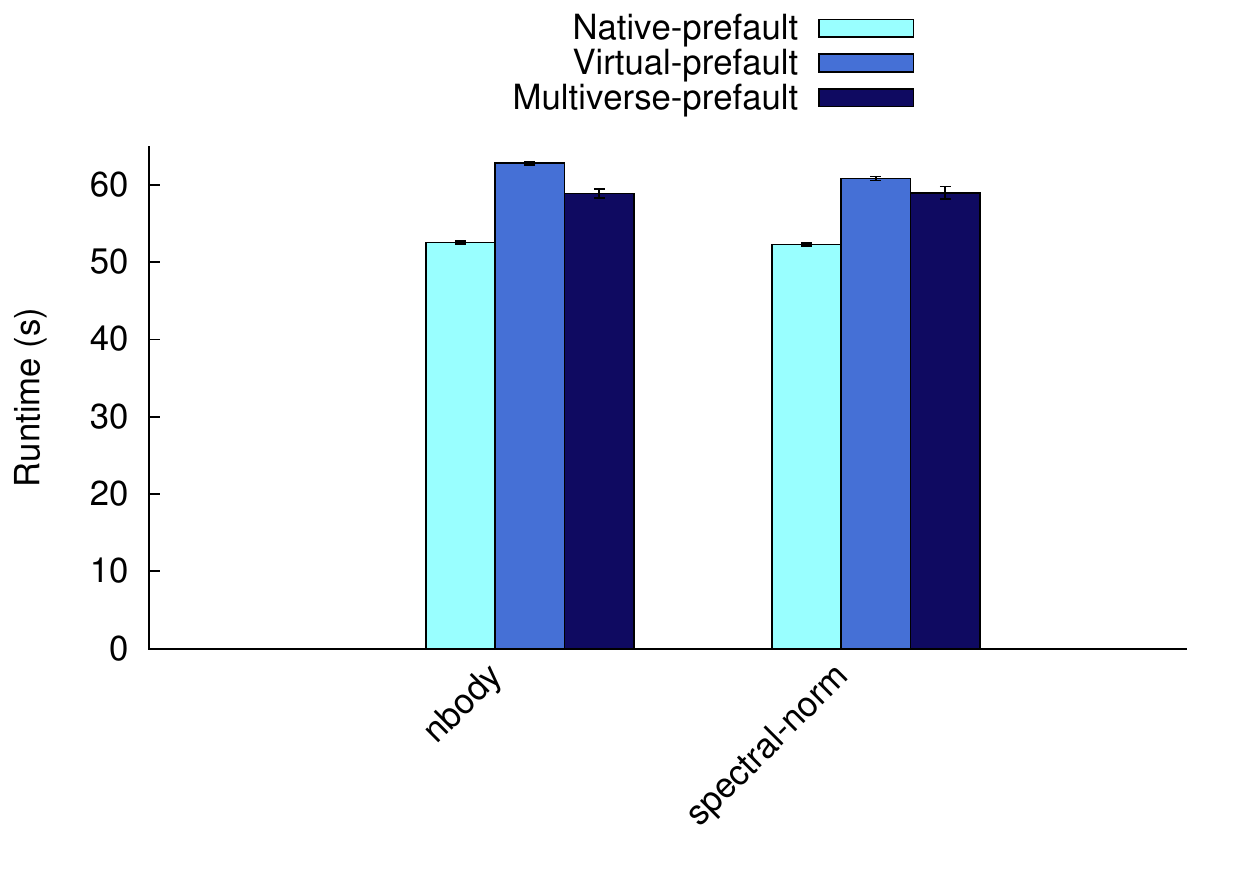}
\cprotect\caption{Performance of nbody and spectral-norm benchmarks with and without a
modification to the runtime that prefaults in all mapped pages, reducing the number of
forwarded events.}
\label{fig:prefault}
\end{figure}

%
%
%
Note that event forwarding will decrease as a runtime is incrementally extended
to take advantage of the HRT model.  As an illustration of this, we modified
Racket's garbage collector to prefault in pages any time that it allocates
a range of memory for internal use. We accomplished this with a simple change
that added the \verb|MAP_POPULATE| flag to Racket's internal \verb|mmap()|
invocations. This reduces the number of page faults incurred during the
execution of a benchmark, therefore reducing the number of forwarded events.
The results are shown in Figure~\ref{fig:prefault}.  While this change
increases the running time for the benchmark overall---indicating that this is
not a change that one would introduce in practice---it shows 
what the relative performance would be with fewer forwarded events.  Indeed,
with fewer events, the hybridized versions (Multiverse-prefault) of these benchmarks running in a VM
now outperform their counterparts (Virtual-prefault).



\begin{framed}
It is worth reflecting on what exactly has happened here: we have
taken a massive (800K line),  complex runtime system off-the-shelf, run it through
Multiverse without changes, and as a result have a version of the
runtime system that correctly runs in kernel mode as an HRT and
behaves identically with virtually identical performance.  To be
clear, {\bf all of the Racket runtime except Linux kernel ABI
  interactions is seamlessly running as a kernel.}  While this
codebase is the endpoint for user-level development, it represents a
{\em starting point} for HRT development in the incremental model.
\end{framed}


\section{Related Work}
\label{sec:related}

As far as we are aware, Multiverse is the only system that {\em automatically}
transforms an existing user-level application into a split execution
environment that runs within the context of both a general-purpose OS
and a specialized OS.

Work on specialized kernels goes back decades, and the design of
Nautilus is heavily influenced by much of this early work, including
Exokernels~\cite{ENGLER:1995:EXTERMINATE,ENGLER:1995:EXOKERNEL},
SPIN~\cite{BERSHAD:1995:SPIN}, Scout~\cite{MONTZ:1995:SCOUT},
KeyKOS~\cite{BOMBERGER:1992:KEYKOS}, and ADEOS~\cite{YAGHMOUR:ADEOS}. This line
of work has recently been revitalized in the context of Unikernels~\cite{MADHAVAPEDDY:2013:UNIKERNELS}.





The Dune system~\cite{BELAY:2012:DUNE} allows a special kernel module to
promote selected processes to ones that can access privileged CPU features on
a legacy Linux system. Dune leverages virtualization support to give
applications the ability to access, for example, page tables and protection
hardware. While Dune gives applications access to previously unavailable
hardware, it does so from within the context of a Linux process. Unlike
Multiverse, Dune does not give the application the capability to run in an
entirely separate OS.


Libra~\cite{AMMONS:2007:LIBRA} bears similarities to our system in its overall
architecture.  A Java Virtual Machine (JVM) runs on top of the Libra libOS,
which in turn executes under virtualization. A general-purpose OS runs in
a {\em controller partition} and accepts requests for legacy functionality from
the JVM/Libra partition. This system involved a manual port. 
In contrast, the HVM gives us a more powerful mechanism for sharing
between the ROS and HRT as they share a large portion of the address space. This
allows us to leverage complex functionality in the ROS like shared libraries
and symbol resolution. Furthermore, the Libra system does not provide a way to
automatically create these specialized JVMs from their legacy counterparts.

The Blue Gene/L series of supercomputer nodes run with a Lightweight Kernel
(LWK) called the Blue Gene/L Run Time Supervisor
(BLRTS)~\cite{ALMASI:2003:BLRTS} that shares an address space with applications
and forwards system calls to a specialized I/O node. While the bridging
mechanism between the nodes is similar, there is no mechanism for porting
a legacy application to BLRTS. Others in the HPC community have proposed
similar solutions that bridge a {\em full-weight} kernel with an LWK in
a hybrid model. Examples of this approach include
mOS~\cite{WISNIEWSKI:2014:MOS}, ARGO~\cite{ARGO}, and
IHK/McKernel~\cite{SHIMOSAWA:2014:IHK}.  The Pisces
Co-Kernel~\cite{OUYANG:2015:COKERNEL} treats performance isolation as its
primary goal and can partition hardware between {\em enclaves}, or
isolated OS/Rs that can involve different specialized OS kernels.  

In contrast to the above systems, the HRT model is the only one that allows
a runtime to act {\em as} a kernel, enjoying full privileged access to the
underlying hardware. Furthermore, as far as we are aware, none of these systems
provide an automated mechanism for producing an initial port to the specialized
OS/R environment.

\section{Conclusions}
\label{sec:conc}

We introduced Multiverse, a system that implements {\em automatic
hybridization} of runtime systems in order to transform them into hybrid
runtimes (HRTs). We illustrated the design and implementation of Multiverse and
described how runtime developers can use it for incremental porting
of runtimes and applications from a legacy OS to a specialized AeroKernel.

To demonstrate its power, we used Multiverse to automatically hybridize the
Racket runtime system, a complex, widely-used, JIT-based runtime. With
automatic hybridization, we can take an existing Linux version of a runtime or
application and automatically transform it into a package that appears
to run just like any other program, but actually executes on
remote cores in kernel-mode, in the context of an HRT, and with full access to
the underlying hardware. We evaluated the performance overheads of an
unoptimized Multiverse hybridization of Racket and showed that performance
varies with the usage of legacy functionality. In cases where such use is
minimized, the hybridized runtime can outperform the baseline. Runtime developers can use
Multiverse to start with a working system and incrementally migrate
hot spot functionality to custom components within an AeroKernel.




\balance
\bibliographystyle{abbrv}
\bibliography{kyle,pdinda}
\end{document}